\documentclass[eps,prd,twocolumn,nofootinbib,showpacs,article]{revtex4}
\usepackage{graphicx,color}
\usepackage[colorlinks=true,linkcolor=blue,citecolor=blue,urlcolor=blue]{hyperref}
\begin{document}

\title{Quantum instability and Ehrenfest time for inverted harmonic oscillator }

\author{Shangyun Wang$^1$}
\email[Corresponding author:]{sywang@hynu.edu.cn}

\author{Songbai Chen$^{2,3}$}
\email[email:]{csb3752@hunnu.edu.cn}

\author{Jiliang Jing$^{2,3}$}
\email[email:]{jljing@hunnu.edu.cn}

\address{$^1$College of Physics and Electronic Engineering, Hengyang Normal University, Hengyang 421002, China}
\address{$^2$Department of Physics, Key Laboratory of Low Dimensional Quantum Structures
and Quantum Control of Ministry of Education, Synergetic Innovation Center for Quantum Effects and Applications, Hunan
Normal University,  Changsha, Hunan 410081, People's Republic of China}
\address{$^3$Center for Gravitation and Cosmology, College of Physical Science and Technology, Yangzhou University, Yangzhou 225009, People's Republic of China}

\begin{abstract}

We used out-of-time order correlators (OTOCs) to investigate the quantum instability and Ehrenfest time for the inverted harmonic oscillator (IHO).
For the initial states located in the stable manifolds of IHO, we find that the corresponding OTOC exhibits identical evolutionary characteristics as the saddle point before the Ehrenfest time.
For the initial states located in the unstable manifolds, the OTOCs still grow exponentially but the time to maintain exponential growth is related to the center position of its wave packet in the phase space.
Moreover, we use the Husimi Q function to visualize the quantum wave packets during the OTOC's exponential growth.
Our results show that the quantum instability exists at arbitrary orbits in the IHO system, and the Ehrenfest time in the IHO system depends not only on the initial system photon number, but also on the central positions of the initial states in the phase space.

\end{abstract}

\maketitle

\section{Introduction}
\label{sec:1}

The correspondence principle proposed by the Copenhagen School of Quantum Mechanics states that the dynamics of quantum systems should converge towards classical dynamics when the quantum number of the system approaches infinity.
The correspondence between classical and quantum Lyapunov exponents in many body quantum chaotic systems is one of the important bases for testing the correspondence principle. The Ehrenfest time $\tau=\frac{1}{\tilde{\lambda}}\ln N$ (where $\tilde{\lambda}$ is the exponential growth rates (EGRs) of OTOC and $N$ is the particle number of system\cite{et1}) gives the scale of time for the center of the wave packet will follow the classical dynamics when this wave packet moves in a smoothly varying potential\cite{et2}. In other words, the correspondence principle remains valid in this time scale.

Recently, the exponential growth behavior of OTOC has been regarded as an indicator of chaos or quantum instability in many quantum
systems\cite{sys1,sys2,sys3,sys4,sys5,sys6,sys7,sys8,sys9,sys10,sys11,sys12,sys13,sys14,sys15,sys16}.
In particular, its growth rates before the Ehrenfest time are closely associated with the classic Lyapunov exponent (CLE) $\lambda$ in the large quantum numbers limit\cite{sys12,sys13,sys14,sys15,sys16,cqlpv1,cqlpv2}.
Therefore, the OTOCs are not only used to diagnose quantum chaos, but also to test the correspondence principle.
In the context of black holes physics, the OTOC can be considered to as an indicator of chaos in gravity dual theory\cite{gr1} and
the corresponding upper bound of the maximum quantum Lyapunov exponent (QLE) is obtained in black hole geometry\cite{et1,sys1,gr2}.
Experimentally, the OTOC has been measured in the ion traps systems\cite{exp1,exp2,exp3,exp4} and in the nuclear magnetic resonance platforms\cite{nmr1,nmr2,nmr3}.

However, an important discovery is that the OTOC in some non-chaotic regular systems also exhibits exponential growth behavior in early time,
and the fast scrambling emerges not only in the chaotic case but also in the regular one\cite{sp,sp1}. In the integrable systems, it is shown that the OTOCs in the vicinity of saddle points could grow exponentially since quantum instability existing at saddle points\cite{sp2}.
This means that the quantum instability is not only possible, but is the rule for generic states in the vicinity of saddle point.
These results are not surprising because the instability of the saddle point causes its CLE to be positive.
The similar behavior of OTOC also appears in the inverted harmonic oscillator system with a Higgs potential\cite{sp3}.
This further indicates that quantum instability in regular systems exists at the saddle points and their surrounding regions.
For orbits which far from the saddle point, quantum instability disappeared.
Then, it is natural to ask whether unstable regular orbits cause the OTOCs exhibit exponential growth behaviors before the Ehrenfest time.

To study the early behaviors of OTOCs which initial states located on unstable regular orbits, we consider the simplest unstable but regular system --- the inverted harmonic
oscillator system, described by the Hamiltonian $\hat{H}=p^2/2m-m\omega^2 q^2/2$ \cite{iho1}. This special system has been widely studied in various aspects. In Ref. \cite{iho0}, the authors based on their description of a complete generalized Airy function-type quantum
wave solution for IHO note that the dynamics of quantum wave packet is dependent upon its initial positions.
Moreover, the saddle point of this regular system has an exponential sensitivity to initial conditions as in the chaotic systems\cite{iho11}.
The phase-space volume of the classical IHO system is unbounded, however, its corresponding volume for quantum inverted oscillator can be bounded by the system photons\cite{ihodicke1,ihodicke2}.
Especially, the IHO system is not just a pure theoretical model and has been realized experimentally\cite{iho2}. In mathematics, it even challenges the Riemann hypothesis\cite{iho3}. Moreover, it also shows important significance in general relativity, quantum mechanics and chaotic domain\cite{sign1,sign2,sign3,sign4,sign5,sign6,sign7}.

Fidelity OTOCs (FOTOCs) are related to the quantum variance of Hermitian operators and provides a method for us to visualize the chaotic or scrambling dynamics of quantum systems in semi-classical phase space.
Recently, its growth rate before the Ehrenfest time was shown to be associated with the CLE of chaotic systems\cite{sys13,sys15} and the quantum instability of saddle point in non-chaotic system \cite{sp2}.
In Schrodinger picture, the evolutions of the quantum wave packets in the phase space are directly related to the quantum variance of momentum or coordinate operator.
On the other hand, Husimi $Q$ function is one of the quasi-probability distribution functions and was used to visualize the evolution of quantum wave packets in phase space\cite{hs1,hs2,EE1,EE2,EE3}.
In this paper, we use OTOCs to study the quantum instability for the IHO system and to see what Husimi quasi-probability wave packets behave as the OTOCs grow exponentially.
In addition, we also discussed the differences of Ehrenfest time between stable and unstable manifolds in IHO system.

This paper is organized as follows. In section II, we study time evolution of mean photon number for different initial states in quantum IHO system.
In section III, we analyze the early behavior of OTOCs for unstable orbits in IHO system and visualize the OTOCs by Husimi Q function.
Finally, we present results and a brief summary.

\section{Mean photon number in inverted harmonic oscillator}
\label{sec:2}

In this section, we study the classical one-dimensional IHO system with unstable regular orbits, namely,
\begin{eqnarray}
H &=& \frac{p^2}{2m} + V,  \ \ \ \ \ \ V = -\frac{1}{2}m\omega^2 q^2,   \label{iho1}
\end{eqnarray}
where $\omega$ and $m$ are the frequency and mass of IHO, respectively.
\begin{figure}[htp]
\begin{center}
\includegraphics[width=6cm]{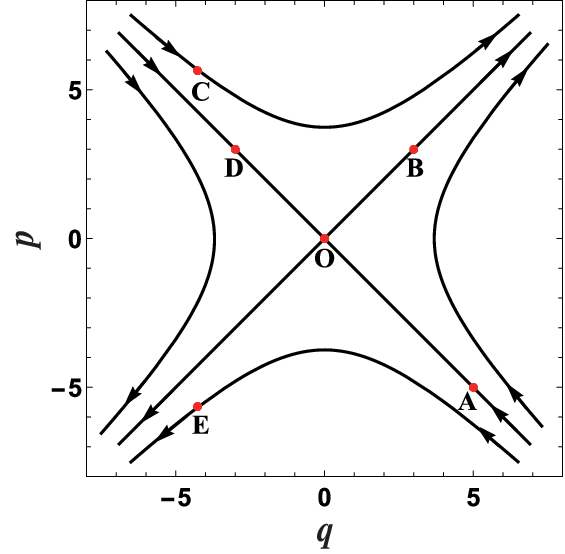}
\caption{The classical dynamics phase diagram of IHO. The coordinate $(q, p)$ of points $O, A, B, C, D$ and $E$ are $(0, 0), (5, -5), (3, 3), (-4.267, 5.643), (-3, 3)$ and $(-4.267, -5.643)$, respectively.
Point $O$ is the saddle point.}
\label{f11}
\end{center}
\end{figure}
In contrast to harmonic oscillator, the classical potential $V$ tends to minus infinity in the limit $q\rightarrow \infty$.
In Fig. \ref{f11}, we present the classical dynamics phase diagram of IHO system.
For the classical Hamiltonian (\ref{iho1}), there is a saddle point $O(q=0, p=0)$ in phase space.
For the particles on the asymptote ($p=-q$), they form a so-called stable manifold since these points move to the saddle point under the action of Hamiltonian (\ref{iho1}).
The particles starting from other positions move to infinity which forms an unstable manifold.
It should be noted that all orbits in the IHO system are unstable, and the Lyapunov exponents are the same as the saddle point, i.e., $\lambda_O=1$. In this paper, we set $m=\omega=1$.

To study the quantum dynamics of IHO, we introduce the quadratic quantization form of the position and momentum operators and set $\hbar=1$,
\begin{eqnarray}
\hat{q} &=& (a^{\dag} + a)/\sqrt{2} ,  \ \ \ \ \ \ \hat{p} = i(a^{\dag} - a)/\sqrt{2},   \label{qq1}
\end{eqnarray}
then the quantum Hamiltonian of IHO becomes to
\begin{eqnarray}
\hat{H} &=& -\frac{1}{2}(a^2 + a^{\dag 2}).   \label{iho2}
\end{eqnarray}

Notice that the Hamiltonian (\ref{iho2}) is equivalent to the Hamiltonian (\ref{iho1}) in the classical limit when the system photon number $N_p\rightarrow \infty$.
As in Refs.\cite{hs2,EE4}, we take the photon coherent state as the initial state of system
\begin{eqnarray}
|\psi\rangle &=& e^{-\beta\beta^{*}/2}e^{\beta a^{\dagger}}|0\rangle, \label{bcoh1}
\end{eqnarray}
with
\begin{eqnarray}
\beta &=& (q + ip)/\sqrt{2}.   \label{tb1}
\end{eqnarray}
where $|0\rangle$ is the ground state of light field, $q$ and $p$ are generalized coordinates and momentum.
According to quantum theory, the mean photon number on the initial coherent state is
\begin{eqnarray}
\langle a^{\dag}a\rangle &=& \langle\psi|a^{\dag}a|\psi\rangle=\beta\beta^{\dag}=(q^2 +p^2)/2.   \label{np1}
\end{eqnarray}
Eq.(\ref{np1}) implies that the mean photon number is related to the central position of coherent state in phase space. In other words,
if the photon number of the system $N_p$ is given, the coordinate and momentum parameters of coherent states in phase space are bounded in the region $(q^2 +p^2)/2\leq N_p$.
This shows that there is a natural boundary in phase space of the quantum IHO system with finite photon number, while the corresponding classical IHO system is unbounded.
\begin{figure}[ht]
\begin{center}
\includegraphics[width=4cm]{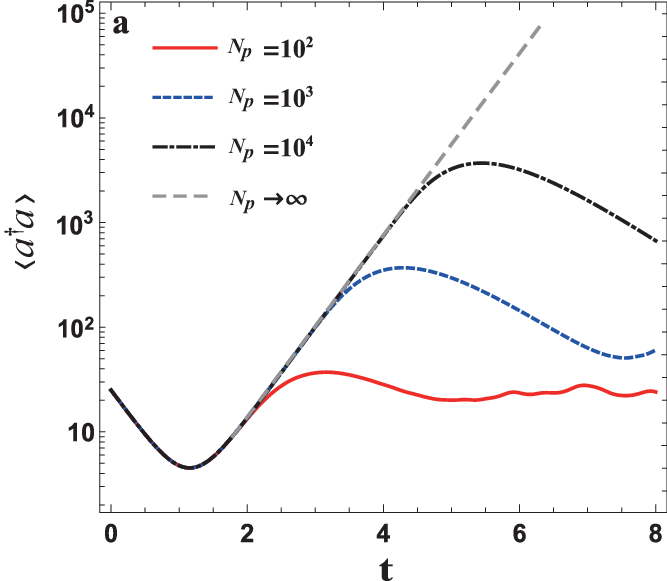}
\includegraphics[width=4cm]{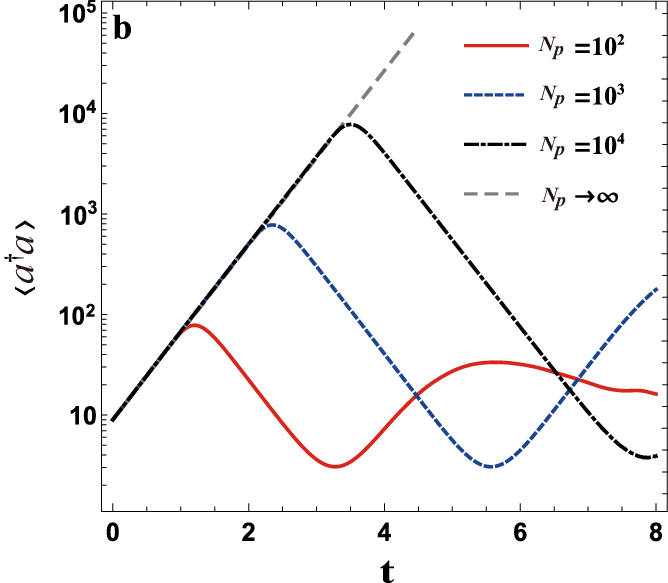}
\caption{Time evolution of the mean photon number for points A (a) and B (b) in Fig. \ref{f11} with different system photon number.}
\label{f2}
\end{center}
\end{figure}

In Fig. \ref{f2}, we show that the time evolutions of mean photon number $\langle a^{\dag}a\rangle$ for initial coherent states centered at points $A$ and $B$ with fixed system photon number.
It's shown that the starting point located in the stable manifold of IHO, such as point $A$, the mean photon number $\langle a^{\dag}a\rangle$ decreases firstly and then increases to the maximum value, as shown in Fig. \ref{f2}(a).
For the points far away from the saddle point under the action of Hamiltonian, such as point $B$, the mean photon number grows to the maximum value directly, as shown in Fig. \ref{f2}(b).
Whatever the center of the initial wave packets are located, the mean photon number $\langle a^{\dag}a\rangle$ increases to infinity in the classical limit case $N_p\rightarrow\infty$.
For the case with finite initial photon number $N_p$, the curve of mean photon number changing with time first overlaps that in the classical limit.
After a time $t_p$, the the evolutionary behaviors of mean photon number are no longer consistent with that in the classical limit case.
This means that during the period $0<t\leq t_p$, there exist the so-called classical-quantum correspondence for the system (\ref{iho1}) because as $t > t_p$ the quantum behavior in the system differs from that in the classical system.
We also note that the time for the mean photon number to remains consistent with the classical case increases with the increase of the system photon number. This indicates that the time $t_p$ increases with the initial photon number $N_p$, which is understandable because the classical-quantum correspondence becomes clearer in the system with larger photon number.
Moreover, the time $t_p$ for maintaining the classical quantum correspondence is less than the time which the mean number of photon increases to its maximum value.
Strictly speaking, the time to maintain the classical quantum correspondence in IHO system is not determined by the time for the mean photon number evolves to its maximum.

Comparing Figures \ref{f2}(a) and \ref{f2}(b), for fixed initial photon number $N_p$, we find that the time to maintain the classical quantum correspondence for the initial coherent state centred at point $A$ is longer than that at point $B$.
It implies that the length of time maintaining classical-quantum correspondence in the quantum IHO depends on the central positions of the initial coherent states of system in the phase space.

\section{Exponential growth of OTOC in inverted harmonic oscillator}
In this section, we study the quantum variance derived from the OTOC in the IHO system and analyse the Husimi quasi-probability wave packet during the OTOC grow exponentially.
The OTOC is defined as\cite{definition}
\begin{eqnarray}
C(t) &=& \langle[\hat{W}(t),\hat{V}(0)]^{\dagger}[\hat{W}(t),\hat{V}(0)]\rangle,   \label{otoc1}
\end{eqnarray}
where $\langle...\rangle$ denotes the expectation values and  $\hat{W}(t)=e^{i\hat{H}t}\hat{W}e^{-i\hat{H}t}$. $\hat{H}$ is a quantum Hamiltonian,
$\hat{W}$ and $\hat{V}$ are two arbitrary local operators.
\begin{figure}[ht]
\begin{center}
\includegraphics[width=4cm]{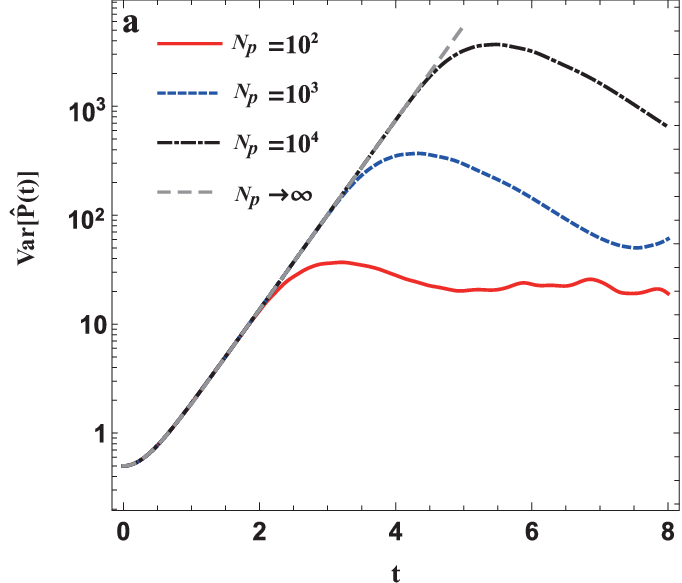}
\includegraphics[width=4cm]{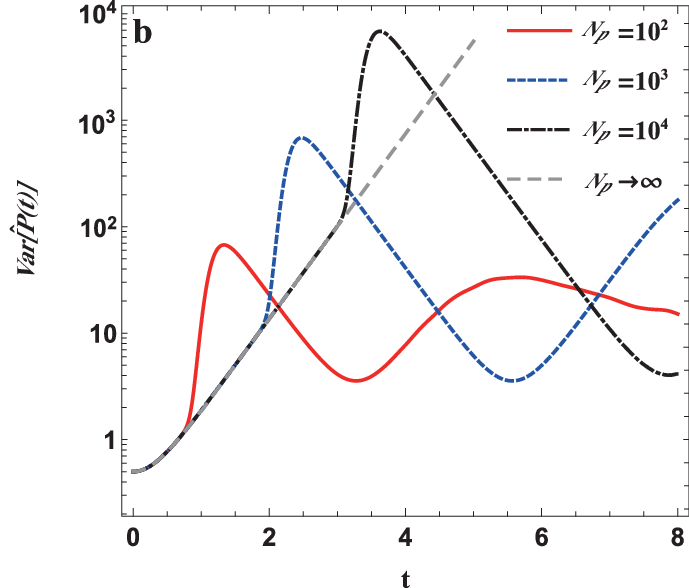}
\caption{Time evolution of the OTOC for point A (a) and B (b) in Fig. \ref{f11} with different system photon number.}
\label{f3}
\end{center}
\end{figure}
Here, we choose $\hat{W}=\hat{P}=i(a^{\dag} - a)/\sqrt{2}$ and $\hat{V}$ as a projection operator onto the initial state $|\psi\rangle$, i.e., $\hat{V}=|\psi\rangle\langle\psi|$.
Substituting the operators $\hat{V}=|\psi\rangle\langle\psi|$ and $\hat{W}=\hat{P}$ into the Eq. (\ref{otoc1}),
one has
\begin{eqnarray}
C(t) &=& \langle\psi(t)|\hat{P}^{2}|\psi(t)\rangle - \langle\psi(t)|\hat{P}|\psi(t)\rangle^2 \\
&\equiv & Var[\hat{P}(t)], \label{otoc3}
\end{eqnarray}
\begin{figure}[ht]
\begin{center}
\includegraphics[width=4cm]{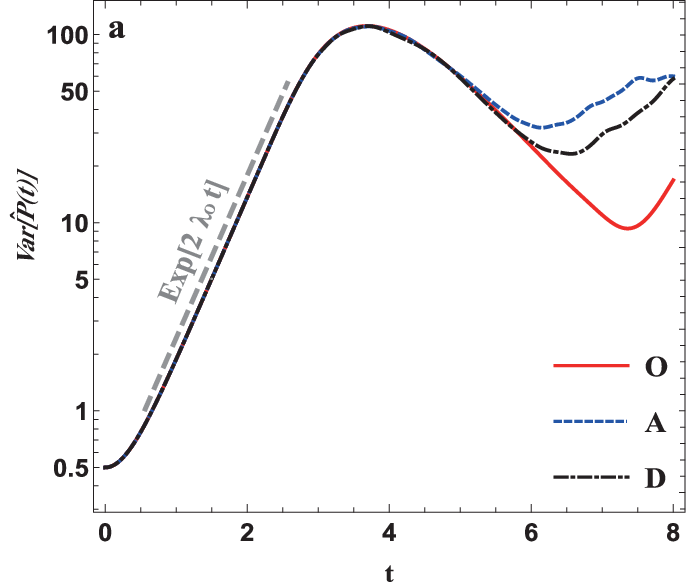}
\includegraphics[width=4cm]{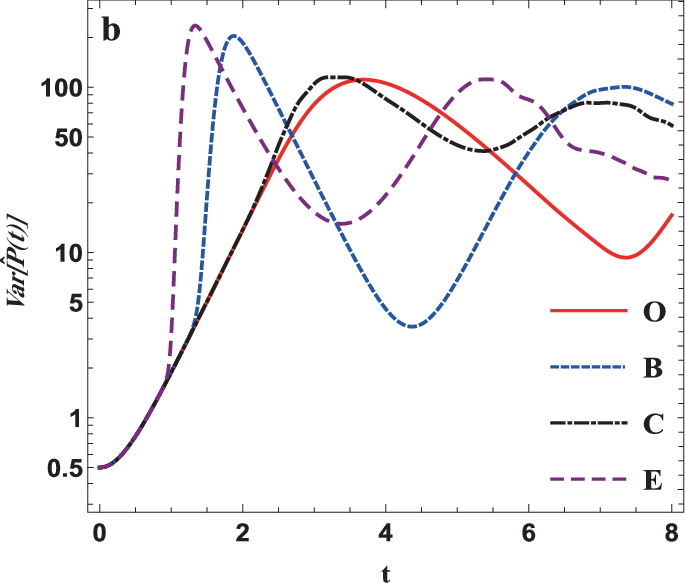}
\caption{Time evolution of the OTOC for the initial coherent states centered at different points. The gray dashed line in the left panel corresponds to the exponential growth rate given by twice the classical Lyapunov exponent. Here, we set the system photon number $N_p = 300$.}
\label{f4}
\end{center}
\end{figure}
where $Var[\hat{P}(t)]$ is the quantum variance of the momentum operator $\hat{P}$.
This relation is similar to the definition of FOTOC\cite{sys13} and enable us to visualize the OTOC in semiclassical phase space.
Whatever the initial states centered at the points A or B, the corresponding OTOCs grow exponentially and the duration of the exponential behavior becomes longer with the increase of system photon number, as clearly seen in Fig. \ref{f3}(a) and Fig. \ref{f3}(b).
Moreover, we find that when the system photon number is fixed, the time for OTOCs to maintain exponential behavior for initial states centered at the point A is longer than that centered at the point B.
Fig. \ref{f3} further indicates that the time to maintain classical-quantum correspondence in the quantum IHO depends not only on the initial system photon number, but also on the central positions of the initial states in the phase space.

\begin{figure*}[htb]
\centering
\includegraphics[width=1.8\columnwidth, angle=0]{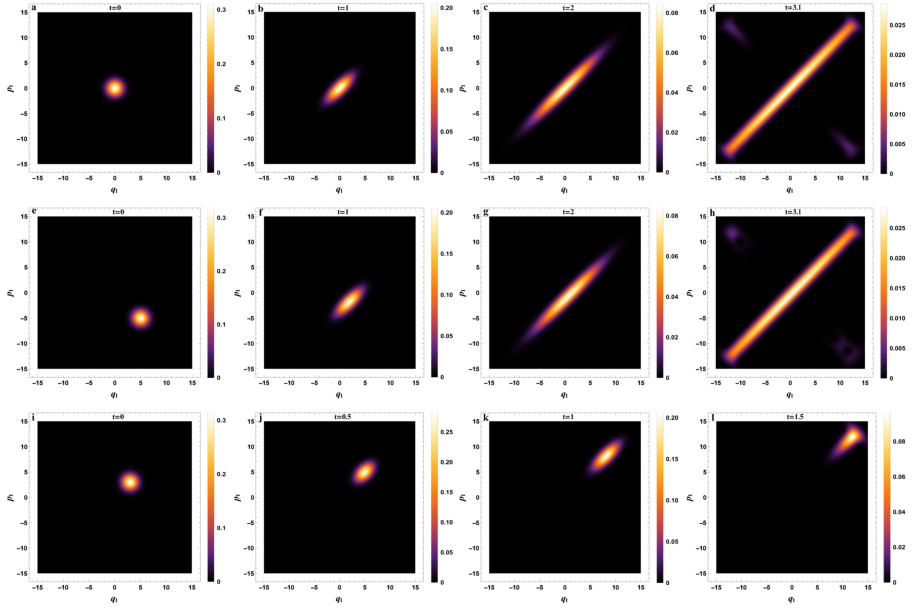}
\caption{Change of Husimi quasi-probabilistic wave packets with time for initial system photon number $N_p = 300$.
The top, middle and bottom panels denote respectively the case in which the initial coherent state centered at the points
$O, A$ and $B$ in Fig. \ref{f11}.}
\label{f5}
\end{figure*}
For the initial states centered at the points in the stable manifold, the evolution of OTOCs are completely consistent before the Ehrenfest time,
and the EGRs are twice the CLE of saddle point, i.e., $\tilde{\lambda}=2\lambda_O$,
as shown in Fig. \ref{f4}(a). Therefore, for the initial states located on the stable manifold, the Ehrenfest time are $\tau=\frac{1}{\tilde{\lambda}}\ln N_p\approx2.8$ when $N_p=300$. In Fig. \ref{f4}(b), we find that the OTOCs with initial states located on the unstable manifold owning the same EGRs as the saddle point, yet the time for OTOCs to maintain exponential growth are related to the position of the initial states. Moreover, comparing Fig. \ref{f4}(a) and \ref{f4}(b), it's evident that the Ehrenfest time for initial states centred in the stable manifold are largest. These results illustrate that the Ehrenfest time in the IHO system depends not only on the initial system photon number, but also on the central positions of the initial states in the phase space.

Husimi Q function is one of the quasi-probability distribution functions, and it was used to distinguish chaotic and periodic orbits in quantum systems\cite{hs1,hs2,EE1,EE2,EE3,EE4}.
The definition of Husimi Q function is
\begin{eqnarray}
Q(q_1,p_1) &=& \frac{1}{\pi}\langle q_1,p_1|\rho|q_1,p_1\rangle,    \label{hq1}
\end{eqnarray}
where $|q_1,p_1\rangle$ is the photon coherent state and $\rho$ is the density matrix. In Fig. \ref{f5}, we present the time evolutions of Husimi Q distribution with different initial states in the phase space of the IHO system.
Figures \ref{f5}(a)-\ref{f5}(c) and \ref{f5}(e)-\ref{f5}(g) show that the evolutions of quasi-probability wave packets for the initial coherent states centered at the points in the stable manifold own similar behaviors during the OTOCs grow exponentially.
This means that the evolution of OTOCs with the initial states centered at the points on the stable manifold are highly consistent before the Ehrenfest time, as shown in Fig. \ref{f4}(a).
In Figs. \ref{f5}(i)-\ref{f5}(l), we show that the evolution of Husimi quasi-probabilistic wave packet with the initial state centered at point $B$ which located on the unstable manifold.
It is found that the time reduces for the wave packet owning the behaviors similar to that on the stable manifold.
During the OTOCs grow exponentially, it is easy to found that in the IHO system, the corresponding quantum wave packets spread along the phase trajectory in the classical phase space.
Especially, we note that the evolution of quasi-probability wave packets are identical except for the position during the OTOCs grow exponentially, as shown in Figs. \ref{f5}(b), \ref{f5}(f) and \ref{f5}(k).
This means that whatever the initial states centered at the points which locate on the stable or unstable manifold, the evolutionary behaviors of OTOCs before Ehrenfest time are consistent, as shown in Fig. \ref{f4}.
In addition, we present the Husimi quasi-probabilistic wave packets after the OTOC grows exponentially in Figs. \ref{f5}(d), \ref{f5}(h) and \ref{f5}(l) for points $O, A$ and $B$, respectively.
For the initial states locate on the stable manifold, when the time exceed the corresponding Ehrenfest time, some discrete wave packets appear in the upper left and lower right corners of the phase space, as seen in Figs. \ref{f5}(d) and \ref{f5}(h).
These evolutionary behaviors of the Husimi Q function are significantly different from that before Ehrenfest time.
This indicates that the classical quantum correspondence have been broken at this time.
For the initial states located on the unstable manifold, the discrete wave packets appear earlier in the upper left and lower right corners of the phase space, as seen in Fig. \ref{f5}(l).
This means that the initial wave packets located in the unstable manifold reach the boundary determined by the initial system photon number more quickly.
Therefore, the classical quantum correspondences in unstable manifold break more earlier and the Ehrenfest time are more shorter than that in stable manifold.
Fig. \ref{f5} provides an intuitive explanation of the behaviors of OTOCs in Fig. \ref{f4} and why the Ehrenfest time in quantum IHO system are related to their initial positions.

\section{Conclusion}
In this paper, we show that both the system photon number and the central position of the initial coherent states affect the Ehrenfest time, and the Ehrenfest time for initial states located on the stable manifold of IHO system are largest.
Moreover, the exponential growth rates of OTOCs at any position in the IHO system are twice the CLE of the saddle point.
This illustrates that quantum instability exists at arbitrarily orbits in the IHO system.
For the starting points located in the stable manifold of IHO, the mean photon number decrease firstly and then increase to the maximum value, and the evolution of OTOCs are completely consistent before the Ehrenfest time.
For the points far away from the saddle point under the action of IHO Hamiltonian, the mean photon number grow to the maximum value directly, and the time for the OTOCs maintaining exponential growth are depend on the central positions of initial wave packets.
Moreover, we analyze the Husimi quasi-probability wave packets of different initial states,
and find that during the OTOCs grow exponentially, the evolution of quasi-probability wave packets are identical except for the position.
Our results could help us to deeply understand correspondence principle and OTOCs in unstable systems.

\section{\bf Acknowledgments}
This work was supported by the National Natural Science Foundation of China under Grant No.12275078,
11875026, 12035005, 2020YFC2201400, and the innovative research group of Hunan Province under Grant No. 2024JJ1006.


\vspace*{0.2cm}
\begin{thebibliography}{99}

\bibitem{et1} J. Maldacena, S.H. Shenker, and D. Stanford, A bound on chaos, J. High Energy Phys. 08 (2016) 106.
\bibitem{et2} Chuan Zhao and Biao Wu, Ehrenfest Time at the Transition from Integrable Motion to Chaotic Motion, Chin. Phys. Lett. 38 030502 (2021).

\bibitem{sys1} S. H. Shenker and D. Stanford, Black holes and the butterfly effect, J. High Energy Phys. 03 (2014) 067.
\bibitem{sys2} D. A. Roberts and D. Stanford, Diagnosing Chaos Using Four-Point Functions in Two-Dimensional
               Conformal Field Theory, Phys. Rev. Lett. 115, 131603 (2015).
\bibitem{sys3} E. R. Castro et al, Quantum-classical correspondence of a system of interacting bosons in a triple-well potential, Quantum 5, 563 (2021).
\bibitem{sys4} M. Rautenberg and M. G\"{a}rttner, Classical and quantum chaos in a three-mode bosonic system, Phys. Rev. A 101, 053604 (2020).
\bibitem{sys5} B. Craps et al, Lyapunov growth in quantum spin chains, Phys. Rev. B 101, 174313 (2020).
\bibitem{sys6} R. K. Shukla, A. Lakshminarayan, and S. K. Mishra, Out-of-time-order correlators of nonlocal block-spin and random observables in integrable and nonintegrable spin chains,
               Phys. Rev. B 105, 224307 (2022).
\bibitem{sys7} M. McGinley, A. Nunnenkamp, and J. Knolle, Slow Growth of Out-of-Time-Order Correlators and Entanglement
               Entropy in Integrable Disordered Systems, Phys. Rev. Lett. 122, 020603 (2019).
\bibitem{sys8} T. Akutagawa, K.Hashimoto, T. Sasaki, and Ryota Watanabe, Out-of-time-order correlator in coupled harmonic oscillators, J. High Energy Phys. 08 (2020) 013.
\bibitem{sys9} E. M. Fortes et al, Gauging classical and quantum integrability through out-of-time-ordered correlators, Phys. Rev. E 100, 042201 (2019).
\bibitem{sys10} R. K. Shukla, A. Lakshminarayan, and S. K. Mishra, Out-of-time-order correlators of nonlocal block-spin and random observables in integrable and nonintegrable spin chains,
                Phys. Rev. B 105 224307 (2022).
\bibitem{sys11} M. McGinley, A. Nunnenkamp, and J. Knolle, Slow Growth of Out-of-Time-Order Correlators and Entanglement
                Entropy in Integrable Disordered Systems, Phys. Rev. Lett. 122, 020603 (2019).
\bibitem{sys12} K. Hashimoto, K. Muratab, and R. Yoshii, Out-of-time-order correlators in quantum mechanics, J. High Energy Phys. 10 (2017) 138.
\bibitem{sys13} R. J. Lewis-Swan, A. Safavi-Naini, J. J. Bollinger, and A. M. Rey, Unifying scrambling, thermalization and entanglement
               through measurement of fidelity out-of-time-order correlators in the Dicke model, Nat. Commun. 10, 1581 (2019).
\bibitem{sys14} J. Ch\'{a}vez-Carlos et al, Quantum and Classical Lyapunov Exponents in Atom-Field Interaction Systems, Phys. Rev. Lett. 122, 024101 (2019).
\bibitem{sys15} A. V. Kirkova, D. Porras, and P. A. Ivanov, Out-of-time-order correlator in the quantum Rabi model, Phys. Rev. A 105, 032444 (2022).
\bibitem{sys16} J. Wang, G. Benenti, G. Casati, and W. Wang, OQuantum chaos and the correspondence principle, Phys. Rev. E 103, L030201 (2021).

\bibitem{cqlpv1} S. H. Shenker and D. Stanford, Black holes and the butterfly effect, J. High Energy Phys. 03 (2014) 067.
\bibitem{cqlpv2} A. Bohrdt, C. B. Mendl, M. Endres, and M. Knap, Scrambling and thermalization in a diffusive quantum many-body system, New J. Phys. 19, 063001 (2017).



\bibitem{gr1} J. M. Maldacena, The Large N limit of superconformal field theories and supergravity, Int. J. Theor. Phys. 38 (1999) 1113.
\bibitem{gr2} S.H. Shenker and D. Stanford, Multiple shocks, J. High Energy Phys. 12 (2014) 046.


\bibitem{exp1} M. G\"{a}rttner, J. G. Bohnet et al, Measuring out-of-time-order correlations
               and multiple quantum spectra in a trapped-ion quantum magnet, Nat. Phys. 13, 781 (2017).
\bibitem{exp2} K. A. Landsman, C. Figgtt et al, Verified quantum information scrambling,
               Nature (London) 567, 61 (2019).
\bibitem{exp3} M. K. Joshi, A. Elben et al, Quantum Information Scrambling in a Trapped-Ion Quantum Simulator with Tunable Range Interactions,
               Phys. Rev. Lett. 124, 240505 (2020).
\bibitem{exp4} A. M. Green, A. Elben et al, Experimental measurement of out-of-time-ordered correlators at finite temperature, Phys. Rev. Lett. 128, 140601 (2022).

\bibitem{nmr1} J. Li, R. Fan, H. Wang, B. Ye, B. Zeng, H. Zhai, X. Peng, and J. Du, Measuring Out-of-Time-Order Correlators on a
               Nuclear Magnetic Resonance Quantum Simulator, Phys. Rev. X 7, 031011 (2017).
\bibitem{nmr2} K. X. Wei, C. Ramanathan, and P. Cappellaro, Exploring Localization
               in Nuclear Spin Chains, Phys. Rev. Lett. 120, 070501 (2018).
\bibitem{nmr3} M. Niknam, L. F. Santos, and D. G. Cory, Sensitivity of quantum information to environment perturbations measured with
               a nonlocal out-of-time-order correlation function, Phys. Rev. Research 2, 013200 (2020).

\bibitem{sp} B. Bhattacharjee, X. Cao, P. Nandy, and T. Pathak . Krylov complexity in saddle-dominated scrambling. J. High Energ. Phys. 2022, 174 (2022).
\bibitem{sp1} T. Xu, T. Scaffidi, and X. Cao, Does Scrambling Equal Chaos?, Phys. Rev. Lett. 124, 140602 (2020).
\bibitem{sp2} Sa\'{u}l Pilatowsky-Cameo et al, Positive quantum Lyapunov exponents in experimental systems with a regular classical limit, Phys. Rev. E 101, 010202(R) (2020).
\bibitem{sp3} K. Hashimoto, K. Huh, K. Kimb, and R. Watanabea, Exponential growth of out-of-time-order correlator
              without chaos: inverted harmonic oscillator, J. High Energy Phys. 11 (2020) 068.

\bibitem{iho1} G. Barton, Quantum mechanics of the inverted oscillator, Ann. Phys. (N.Y.) 166, 322 (1986).
\bibitem{iho0} M. Maamache and J. R. Choi, Quantum-classical correspondence for the inverted oscillator, Chinese Physics C, 2017, 41(11): 113106.

\bibitem{iho11} R. Blume-Kohout and W. Zurek, Decoherence from a chaotic environment: An upside-down oscillator as a model, Phys. Rev. A 68, 032104 (2003).

\bibitem{ihodicke1} K. Gietka, T. Busch, Inverted harmonic oscillator dynamics of the nonequilibrium phase transition in the Dicke model, Phys. Rev. E 104, 034132 (2021).
\bibitem{ihodicke2} K. Gietka, Squeezing by critical speeding up: Applications in quantum metrology, Phys. Rev. A 105, 042620 (2022).
\bibitem{iho2} S. Gentilini, M. C. Braidotti, G. Marcucci, E. DelRe, and C. Conti, Physical realization of the Glauber quantum oscillator, Sci. Rep. 5, 15816 (2015).
\bibitem{iho3} M. V. Berry and J. P. Keating, The Riemann zeros and eigenvalue asymptotics, SIAM Rev. 41, 236 (1999).

\bibitem{sign1} S. Choudhury et al, Four-mode squeezed states in de Sitter space: A study with two field interacting quantum system, arXiv:2203.15815.
\bibitem{sign2} L. Qu, J. Chen, Y. Liu, Chaos and Complexity for Inverted Harmonic Oscillators, Phys. Rev. D 105, 126015 (2022).
\bibitem{sign3} V. Subramanyan et al, Physics of the Inverted Harmonic Oscillator: From the lowest Landau level to event horizons, Annals of Physics 435 (2021) 168470.
\bibitem{sign4} Z. Tian et al, Verifying the upper bound on the speed of scrambling with the analogue Hawking radiation of trapped ions, Eur. Phys. J. C 82, 212 (2022).
\bibitem{sign5} Z. Lewis and T. Takeuchi, Position and Momentum Uncertainties of the Normal and Inverted Harmonic Oscillators under the Minimal Length Uncertainty Relation, Phys. Rev. D 84, 105029 (2011).
\bibitem{sign6} P. Betzios, N. Gaddam, and O. Papadoulaki, Black holes, quantum chaos, and the Riemann hypothesis, SciPost Phys. Core 4, 032 (2021).
\bibitem{sign7} T. Morita, Thermal Emission from Semiclassical Dynamical Systems, Phys. Rev. Lett. 122, 101603 (2019).

\bibitem{hs1} K. Takahashi and N. Sait\^{o}, Chaos and Husimi Distribution Function in Quantum Mechanics, Phys. Rev. Lett. 55, 645 (1985).
\bibitem{hs2} S. Wang, S. Chen and J. Jing, Effect of system energy on quantum signatures of chaos in the two-photon Dicke model, Phys. Rev. E 100, 022207 (2019).
\bibitem{EE1} S. Chaudhury, A. Smith, B. E. Anderson, S. Ghose and P. S. Jessen, Quantum signatures of chaos in a kicked top, Nature (London) 461, 768 (2009).
\bibitem{EE2} A. Piga, M. Lewenstein and J. Q. Quach, Quantum chaos and entanglement in ergodic and nonergodic systems, Phys. Rev. E 99, 032213 (2019).
\bibitem{EE3} V. Mourik et al, Exploring quantum chaos with a single nuclear spin, Phys. Rev. E 98, 042206 (2018).
\bibitem{EE4} K. Furuya, M. C. Nemes, and G. Q. Pellegrino, Quantum Dynamical Manifestation of Chaotic Behavior in the Process of Entanglement,
              Phys. Rev. Lett. 80, 5534(1998).

\bibitem{sm1} See Supplemental Material at http://link.aps.org/supplemental/10.1103/PhysRevE.10
1.010202 for details.

\bibitem{definition} A. I. Larkin and Yu. N. Ovchinnikov, Zh. Eksp. Teor. Fiz. 55, 2262 (1969) [Sov. Phys. JETP 28, 1200 (1969)].

\bibitem{tsm} J. Ch$\acute{a}$vez-Carlos, Classical chaos in atom-field systems. Phys. Rev. E 94, 022209 (2016).

\end{thebibliography}
\end{document}